\documentclass[prl,twocolumn,10pt,a4paper,superscriptaddress,showpacs]{revtex4}
\usepackage{amsmath,graphicx,amsfonts,epstopdf}

\def\opone{\leavevmode\hbox{\small1\kern-3.8pt\normalsize1}}

\begin{document}

\title{Analytical Solution of Cross Polarization Dynamics}
\author{Peng Li}
\author{Qun Chen}
\email{qchen@admin.ecnu.edu.cn}
\author{Shanmin Zhang}
\email{shanminz@hotmail.com}
\affiliation{Department of Physics and Shanghai Key Laboratory of Magnetic
Resonance, East China Normal University, 3663 North Zhongshan Road,
Shanghai 200062, P. R. China}

\begin{abstract}
Cross polarization (CP) dynamics, which was remained unknown for five
decades,  has been derived analytically in the zero- and double-quantum
spaces.  The initial polarization in the double-quantum space is a constant
of motion under strong pulse condition ($|\omega_{1I}+\omega_{1S}|\gg |d(t)|$),
while the Hamiltonian in the zero-quantum space reduces to $d(t)\sigma_{z}^{\Delta}$ under the Hartmann-Hahn
match condition ($\omega_{1I}=\omega_{1S}$).  The time dependent Hamilontian ($d(t)\sigma_{z}^{\Delta}$) in the zero-quantum
space can be expressed by average Hamiltonians.  Since$[d(t')\sigma_{z}^{\Delta},
d(t'')\sigma_{z}^{\Delta}]=0$, only zero order average Hamiltonian needs
to be calculated, leading to an analytical solution of CP dynamics.
\end{abstract}

\maketitle

\section{1. Introduction}

Cross polarization (CP), developed by Hartmann and Hahn~\cite{HartmannHah} in 1962 and later modified
by Pines et al~\cite{Pines}, is one of the most important techniques in NMR.  Mediated by
heteronuclear dipolar interaction, polarization is often transferred from  abundant
I spins to a rare S spin with a polarization enhancement up to $\gamma_{I}/\gamma_{S}$. Because the overall delay time
is just only subject to the spin-lattice relaxation time of I spins, considerable NMR time is saved compared with a single pulse experiment.

CP dynamics for a single crystal was first derived by M\"{u}ller et al~\cite{Ernst}. For a stationary sample,
the Hamiltonian is time independent, it can also be solved by diagonalization of the Hamiltonians. Under fast magic angle spinning(MAS)
($\omega_r/2\pi\gg|d(t)|$), the Hartman-Hahn match condition shift to $\omega_{1I}-\omega_{1S}=n\omega_r$($n=\pm1,2$)~\cite{Stejskcal}
and the effective heteronuclear dipolar interaction becomes time independent~\cite{Shanmin1}. Consequently the CP dynamics under fast MAS
can be derived without many problems~\cite{Shanmin1}, in particular under Lee-Goldburg (LG)~\cite{LG} condition as demonstrated by Ladizhansky et al~\cite{LGCP, Vega}.
However, under conventional MAS speed, the heteronuclear dipolar interaction becomes time dependent, as many other quantum systems,
searching for an analytic solution is usually not conceivable.

So far, CP dynamics under conventional MAS speed is usually calculated with a spin temperature hypothesis~\cite{Mehring,Goldman}.  It
inevitably leads to an empirical solution~\cite{ChemRev}.  In the zero-
and double-quantum spaces~\cite{Levitt}, which commutes with each other, the evolution of density matrix can be done separately in the two spaces,
simplifying the calculation considerably. By means of this method, a number of intriguing phenomena in CP has been
understood thoroughly.  They include CP dynamics of phase-shifted CP under
mismatch conditions~\cite{Levitt, Shanmin2}, W-MOIST~\cite{WMOIST}, double-quantum matched CP~\cite{Meier,Shanmin1},
and adiabatic polarization transfer~\cite{Adiabatic, Shanmin1}.

Up to now, CP dynamics under conventional MAS speed and the Hartman-Hahn match condition ($\omega_{1I}=\omega_{1S}$) has
remained unknown.  In this article we show that this problem can be resolved in the zero- and
double-quantum spaces together with average Hamiltionian theory~\cite{AHT}.  The experiment and simulated results match well with the theoretical predictions.

\section{2. Theory}

For a heteronuclear dipolar coupled $IS$ spin pair under cross polarization
and magic angle spinning (CPMAS), the Hamiltonian can be described by
\begin{equation}
\mathcal{H}=\omega_{1I}I_{y}+\omega_{1S}S_{y}+2d(t)I_{z}S_{z}, \label{totalHamiltonian}
\end{equation}
where $\omega_{1I}$ and $\omega_{1S}$  are the strengths of the I and S spin-locking
fields, respectively, and
\begin{equation}
\begin{split}
d(t)&=d[\sqrt{2}sin(2\beta)cos(\omega_{r}t+\gamma)\\
&-sin^{2}(\beta)cos(2\omega_{r}t+2\gamma)]
\end{split}
\end{equation}
is a time dependent heteronuclear dipolar coupling constant with two Euler angles in
the rotor-fixed frame. In the zero- and double-quantum spaces, the above Hamiltonian
can be expressed as~\cite{Levitt,Shanmin1,Shanmin2,WMOIST}
\begin{equation}
\mathcal{H}^{\Delta}=d(t)\sigma_{z}^{\Delta}+(\omega_{1I}-\omega_{1S})\sigma_{y}^{\Delta}\ (zero-quantum) \label{zeroQuantum}
\end{equation}
and
\begin{equation}
\mathcal{H}^{\Sigma}=d(t)\sigma_{z}^{\Sigma}+(\omega_{1I}+\omega_{1S})\sigma_{y}^{\Sigma}\ (double-quantum), \label{doubleQuantum}
\end{equation}
where $\sigma_{i}^{\Delta}$ and $\sigma_{i}^{\Sigma}$ are the Pauli matrices.
The initial density matrix can be expressed in terms of zero-quantum and double-quantum density matrices
\begin{equation}
\rho(0)=I_{y}=\frac{1}{2}(I_{y}-S_{y})+\frac{1}{2}(I_{y}+S_{y})=\sigma_{y}^{\Delta}+\sigma_{y}^{\Sigma}. \label{initialState}
\end{equation}
Because $[\sigma_{i}^{\Delta}, \sigma_{i}^{\Sigma}]=0(i,j=x,y,z)$
and $[\mathcal{H}^{\Delta}, \mathcal{H}^{\Sigma}]=0$,
the evolution of density matrix can be calculated in the zero- and double-quantum spaces separately,
\begin{equation}
\begin{split}
\rho(t)&=\sigma^{\Delta}(t)+\sigma^{\Sigma}(t)
\\
&=e^{-i\int_{0}^{t}\mathcal{H}^{\Delta}(t')dt'}\sigma_{y}^{\Delta}e^{\int_{0}^{t}\mathcal{H}^{\Delta}(t')dt'}
\\
&+e^{-i\int_{0}^{t}\mathcal{H}^{\Sigma}(t')dt'}\sigma_{y}^{\Sigma}e^{\int_{0}^{t}\mathcal{H}^{\Sigma}(t')dt'}. \label{evolution}
\end{split}
\end{equation}
Under strong pulse condition $|\omega_{1I}+\omega_{1S}|\gg |d(t)|$,
the density matrix in the double-quantum space is nearly a
constant of motion, i.e.
\begin{equation} \label{evo_double}
\sigma^{\Sigma}(t)\approx\sigma_{y}^{\Sigma}.
\end{equation}
Therefore, the calculation of density matrix is largely determined by the evolution in
the zero-quantum space.  Under Hartmann-Hahn match condition $\omega_{1I}=\omega_{1S}$,
the Hamiltonian in the zero-quantum space becomes
\begin{equation}
\mathcal{H}^{\Delta}(t)=d(t)\sigma_{z}^{\Delta},
\end{equation}
which is time dependent.  For every particular time,
it can be represented by a zero-order average Hamiltonian
\begin{equation}\label{zeroAHT}
\begin{split}
\overline{\mathcal{H}}^{\Delta(0)}&(t)=\frac{\sigma_{z}^{\Delta}}{t}\int_{0}^{t}d(t')dt'=\overline{d}(t)\sigma_{z}^{\Delta}
\\
&=\frac{d}{2t}\{2\sqrt{2}sin(2\beta)[sin(\omega_{r}t+\gamma)-sin(\gamma)]
\\
&-sin^{2}(\beta)[sin(2\omega_{r}t+2\gamma)-sin(2\gamma)]\}\sigma_{z}^{\Delta}.
\end{split}
\end{equation}
Since $[\mathcal{H}^{\Delta}(t'), \mathcal{H}^{\Delta}(t'')]=0$ for any $t'$ and $t''$,
all the higher order Hamiltonians vanish.  Consequently, the zero-order average
Hamiltonian represents exactly the Hamiltonian of the spin system.  It turns out
that the average Hamiltonian itself is time dependent, but for any particular time t
it can be treated as a time independent Hamiltonian in the evolution of density matrix.
Therefore, for a given average Hamiltonian the density matrix in the zero-quantum space can be derived
\begin{equation} \label{evo_zero}
\begin{split}
\sigma^{\Delta}(t)&=e^{-i\overline{d}(t)\sigma_{z}^\Delta}\sigma_{y}^{\Delta}e^{i\overline{d}(t)\sigma_{z}^\Delta}\\
&=cos(\overline{d}(t))\sigma_{y}^{\Delta}-sin(\overline{d}(t))\sigma_{x}^{\Delta}.
\end{split}
\end{equation}
Base on the Eqs. (\ref{evolution}), (\ref{evo_double}), and (\ref{evo_zero}), the evolution of density matrix becomes
\begin{equation}
\begin{split}
\rho(t)&=\sigma^{\Delta}(t)+\sigma^{\Sigma}(t)
\\
&=cos(\overline{d}(t))\sigma_{y}^{\Delta}-sin(\overline{d}(t))\sigma_{x}^{\Delta}+\sigma_{y}^{\Sigma},
\end{split}
\end{equation}
which leads to a CP dynamics
\begin{equation} \label{CPD}
\begin{split}
CP(t) &= Trace<S_{y} \cdot \rho(t)>\\
&=Trace<(\sigma_{y}^{\Sigma}-\sigma_{y}^{\Delta})\cdot \rho(t)>\\
&=1-cos(\overline{d}(t)).
\end{split}
\end{equation}
The CP dynamics of powder sample can be derived by integration over the Euler solid angle on the sphere
\begin{equation} \label{powder}
CP(t)=\int_{0}^{\pi}\!\!\!\int_{0}^{2\pi}\!\!\!\{1-cos[\overline{d}(t)]\}sin(\beta)d\beta d\gamma.
\end{equation}
Considering the effects of I spin diffusion and spin lattice relaxation in the rotating frame, the CP dynamics becomes
\begin{equation} \label{diffusion}
\begin{split}
M(t)&=M_{0}\cdot\{1-\frac{1}{2}e^{-Rt}\\
&-\frac{1}{2}e^{-R_{1}t}(1-2\cdot CP(t))\}\cdot e^{-\frac{t}{T_{1\rho}}},
\end{split}
\end{equation}
where $R$ is the spin diffusion rate of I spin, $R_{1}$ is a rate which results in the oscillation damping by the remote
I spins, and $T_{1\rho}$ is the spin lattice relaxation in the rotating frame of the I spins.

When $\omega_{r}\rightarrow 0$, the Eq. (\ref{diffusion}) reduce to:
\begin{equation} \label{crystal}
M(t)=M_{0}\cdot\{1-\frac{1}{2}e^{-Rt}-\frac{1}{2}e^{-R_{1}t}cos(dt)\}\cdot e^{-\frac{t}{T_{1\rho}}}
\end{equation}
which, except for the T1 effect, is identical to the result by M\"{u}ller et al~\cite{Ernst} for a stationary sample.

In the above calculation, the offsets, $\Delta\omega_{I}$ and $\Delta\omega_{S}$ for I and S spin, are ignored.  
Under off-resonance condition, the Hartman-Hahn match condition is determined by effective fields, i.e. $\omega_{1Ie}=\omega_{1Se}$, where
\begin{equation} \label{effective_field}
\begin{split}
&\omega_{1Ie}=\sqrt{\Delta\omega_{I}^2+\omega_{1I}^2} \qquad and
\\
&\omega_{1Se}=\sqrt{\Delta\omega_{S}^2+\omega_{1S}^2}.
\end{split}
\end{equation}
Correspondingly, the dipolar interaction is split up into two parts (perpendicular and parallel) 
\begin{equation} \label{effective_dipole}
\begin{split}
d(t)I_{z}S_{z} &= sin(\theta_{I})sin(\theta_{S})d(t)I_{ze}S_{ze} \\
&+cos(\theta_{I})cos(\theta_{S})d(t)I_{ye}S_{ye},
\end{split}
\end{equation}
where
\begin{equation} \label{effective_angle}
\begin{split}
&\theta_{I}=cos^{-1}(\Delta\omega_{I}/\omega_{1Ie}) \qquad and \\
&\theta_{S}=cos^{-1}(\Delta\omega_{S}/\omega_{1Se}).
\end{split}
\end{equation}
In Eq. (\ref{effective_dipole}), the perpendicular term is responsible for polarization transfer while the parallel term is a small perturbation which can be ignored here.
In the perpendicular term, the dipolar interaction appears to be scaled by $sin(\theta_{I})sin(\theta_{S})$.  It in turn cases a scaled polarization transfer rate by the 
same factor. In this case, the initial polarization should be placed
along the effective filed of I spin, and the polarization of S spin is built up along the effective field of S spin.

\section{3. Computer Simulation}

\begin{figure}
 \includegraphics[width=0.8\columnwidth]{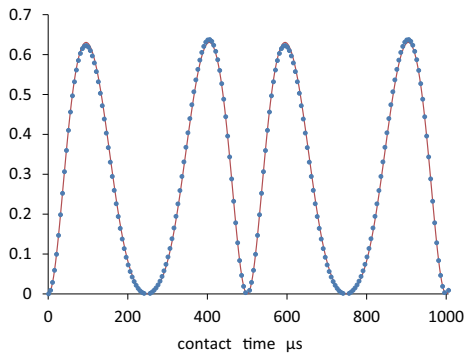}\\
 \caption{Comparison of CP dynamics derived from the analytical solution (solid line) and QCNMR (solid circles) with the conditions:
  $\omega_{1I}/2\pi=\omega_{1S}/2\pi$=80kHz and heteronuclear dipolar constant $d/\pi$=5kHz.
}
\label{figure1}
\end{figure}

\begin{figure}
 \includegraphics[width=0.8\columnwidth]{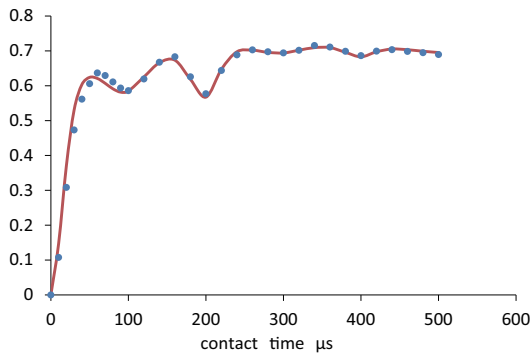}\\
 \caption{CP dynamics from a powder sample of alanine derived from the analytical solution (solid line) and experiments with the conditions:
 $R^{-1}=290.8\mu sec,\ R_{1}^{-1}=137.9\mu sec,\ T_{1\rho}=1.867 msec,\ \omega_{1I}/2\pi=\omega_{1S}/2\pi$=80kHz, and the distance between $^{1}$H and $^{13}$C spin is 1.09\AA~\cite{CH_distance}. The experiment was done by Bruker Avance 300MHz NMR Instrument.
}
\label{figure2}
\end{figure}

A home-made Java program termed QCNMR (quantum computation of NMR),which is based on the evolution of density matrix, is
used for comparing the results from the analytical solution and computer simulation.
All the solid lines shown in FIG.\ref{figure1} and FIG.\ref{figure2} are derived from the analytic solution (Eq. (\ref{CPD}) and (\ref{diffusion})) while the solid
circles in the figures are given by the computer simulation and NMR experiment, respectively.

The CP dynamics with a MAS speed of 2 kHz is shown in FIG.\ref{figure1}.  It appears to be periodic with
a period which is the same as the period of MAS ($T=2\pi/\omega_{r}$).  The polarization of initial buildup
and two nulls within the period ($T=500us$) are caused by the interference between dipolar
oscillation and MAS. This pattern is unique in slow MAS speed. In the above calculation
$R,\ R_{1}$, and $T_{1\rho}$ in the equation are all neglected for better comparison between
theory and simulation.

In FIG.\ref{figure2}, we show the dynamics of a powder sample under a MAS speed of 5 kHz.  Unlike a single
crystal, the oscillation is strongly damped by the orientations of heteronuclear dipolar
tensors.  The polarization increases gradually as the CP and spin diffusion take place. The experiment
results are normalized according to quantitative CP experiment with a reciprocity
relation~\cite{Shujie, Shuwenfang}, while the solid line is normalized by the Eq. (\ref{diffusion}).
In this case the match of two curves depends not only on the patterns but also on the specific values as well.

It can be seen from FIG.\ref{figure1} and FIG.\ref{figure2}, all the simulated results agree well with the
analytical solutions, demonstrating the validity of the analytical solution.  The only approximation in
the derivation is $|\omega_{1I}+\omega_{1S}|\gg |d(t)|$, which is well fulfilled in practice.

\section{4. Conclusion}

Under Hartman-Hahn match condition, CP dynamics can be derived analytically in
the zero- and double-quantum spaces.  The solution is valid for any
MAS speed and offsets.  In particular, the dynamics for a stationary sample
appears when MAS speed approaches zero.  As many other methods,
the CP dynamics provides valuable molecular structural information.
Similar to REDO experiment~\cite{REDO}, this analytic solution also provides a
measure of dipolar coupling constant(or distance) for a strongly coupled system
that is surrounded by a moderately coupled network. Unlike many other
others ~\cite{Vega, MeiHong, F_Si, C_H, H_P, F_C}, this method does not required high MAS speed. If LG spin locking
is applied to proton channel, the result of Least-Square fitting in FIG.\ref{figure2} will
be better because all homonuclear coupling is decoupled.

For a time dependent system, it is unlikely to find a systematic way for analytic solutions. This
may explain why CP dynamics discussed in this article was delayed for so long a time. For an
inhomogeneously broaden system ($[\mathcal{H}(t'), \mathcal{H}(t'')]=0$)~\cite{Broaden}, all high order
average Hamiltonians become zero except for the zero order average Hamiltonian that can be calculate
conveniently. This method is quite general and can be used in NMR, optics, quantum computing
and quantum mechanic related problems of a similar nature.

\section{acknowledgment}

This work is supported by National Fundamental Research Project of China (2007CB925200). Peng Li is grateful "PhD Program Scholarship Fund of ECNU 2007".  Qun Chen is grateful for "Shanghai Leading Talent Training Program" and the support from Shanghai Committee of Science and Technology (11JC1403600).

\end{document}